# Effect of impurity phase and high-pressure synthesis on the superconducting properties of CaKFe$_4$As$_4$

Manasa Manasa[1], Mohammad Azam[1], Tatiana Zajarniuk[2], Ryszard Diduszko[3], Tomasz Cetner[1], Andrzej Morawski[1], Andrzej Wiśniewski[2], Shiv J. Singh[1*]

[1]*Institute of High Pressure Physics (IHPP), Polish Academy of Sciences, Sokołowska 29/37, 01-142 Warsaw, Poland*

[2]*Institute of Physics, Polish Academy of Sciences, aleja Lotników 32/46, 02-668 Warsaw, Poland*

[3]*Łukasiewicz Research Network Institute of Microelectronics and Photonics, Aleja Lotników 32/46, 02-668 Warsaw, Poland*

*Corresponding author:

Email: sjs@unipress.waw.pl

# Abstract

$AeA$Fe$_4$As$_4$ ($Ae$ = Ca, $A$ = K; 1144) having a transition temperature of 35 K is a stoichiometric family of iron-based superconductors (FBS). Here, we present a detailed study of a high-pressure synthesis of CaKFe$_4$As$_4$ bulks to investigate the impact of these conditions on the superconducting properties of the 1144 family. Additionally, these samples are also prepared by conventional synthesis method at ambient pressure (CSP) and studied the influence of impurities on the superconducting properties of CaKFe$_4$As$_4$. Structural, microstructural, transport and magnetization measurements have been performed to reach the final conclusions. Interestingly, the high-pressure synthesis of the parent CaKFe$_4$As$_4$ compound enhances the transition temperature ($T_c$) by 2 K and the critical current density ($J_c$) by one order of magnitude in the whole magnetic field range of 9 T than that of the 1144 bulks prepared by CSP. It suggests the improvement of the pinning centers and grain connections by the high-pressure synthesis approach. Interestingly, our results depict the superconducting onset transition temperature ($T_c^{onset}$) of this stoichiometric 1144 family is robust with the presence of the common 122 (CaFe$_2$As$_2$ or KFe$_2$As$_2$) impurity phases, which is a different behavior compared to other FBS families. However, these impurity phases reduce the grain connections and lead to the phase separation during the formation of the superconducting (1144) phase.

# Introduction

Iron-based superconductors (FBS) [1] are a promising candidate for practical applications because of their high critical current density ($J_c$) of $10^7$-$10^8$ A/cm$^2$ [2], high upper critical field ($H_{c2}$) of 100 T [3, 4], and high transition temperature ($T_c$) of 58 K [5]. Generally, the parent compound does not depict the superconducting properties but shows a structural and magnetic transition around 150 K, and the superconductivity can appear with suitable doping [6, 7]. The crystal structure of the parent compound allows for the classification of over 100 compounds belonging to this high $T_c$ family into six families: *RE*FeAsO (*RE*1111; *RE* = rare earth), 1144 (*AeA*Fe$_4$As$_4$; *Ae* = Ca; *A* = K), *A*Fe$_2$As$_2$ (122; *A* = Ba, K, Ca), (Li/Na)FeAs (111), thick perovskite-type oxide blocking layers, such as Sr$_4$V$_2$O$_6$Fe$_2$As$_2$ (22456), Sr$_4$Sc$_2$O$_6$Fe$_2$P$_2$ (42622), etc., and chalcogenide Fe*X* representing 11 (*X* = chalcogenide). As a doped system, the 1111 family has been reported to have the highest $T_c$ of 58 K [5], while the 1144 family, as a stoichiometric family, has been reported to have the highest $T_c$ of 36 K [8]. It is well known that controlling doping levels in FBS during the growth process is not an easy task, and because of this, there is always a challenge to grow the high-quality samples [7]. Nevertheless, the stoichiometric family, such as 1144, has depicted the superconductivity without any doping issues [9, 10, 11] and the critical current density of the order of $10^8$ A/cm$^2$ (0 T and 5 K) for the single crystal CaKFe$_4$As$_4$, which is the highest $J_c$ value for this high $T_c$ FBS [2]. Hence, this family is highly intriguing for basic and applied research without considering an impurity or disorder effect due to doping effects [12].

The 1144 family has a layered structure of either alkaline or alkaline-earth metals intercalated between the superconducting Fe-As planes [8, 13, 14]. The growth of the 1144 sample has a very narrow synthesis temperature window where the slightly higher or lower temperature promotes the formation of the CaFe$_2$As$_2$ and KFe$_2$As$_2$ (122) phases as an impurity [14, 11]. The formation of these impurity phases generally reduces the sample quality and superconducting properties, as previously reported [11, 10, 9]. The preparation of 1144 bulks using spark plasma sintering (SPS) at a single pressure of 50 MPa revealed a $T_c$ value of 35 K and a critical current density of the order of $10^4$ A/cm$^2$ with a sample density of 96% [9]. Still, there were numerous phases of impurities in this bulk sample. In the SPS technique, the sample is in direct contact with the conducting die or another internal component of the technique that can introduce impurities into the sample. Additionally, the capacity of the sintering sample chamber is constrained by the dimensions and quantity of the samples. Furthermore, the mechanochemical growth process was also performed for CaKFe$_4$As$_4$ (1144) and suggests a slightly lower synthesis temperature, *i.e.* 700°C [10], but still, there is a challenge for the purification and densification of the 1144 samples. The phase pure and highly dense 1144 polycrystalline samples can be produced by the conventional synthesis process at ambient pressure (CSP) and the optimal synthesis conditions (~955°C, 6 h), according to the previous studies [11]. However, their grain connectivity is still very poor, which generally play an important role for the transport properties, particularly for the critical current properties and the application of these materials such as superconducting wires [15]. As a result of these factors, the critical current density of the 1144 polycrystalline samples is significantly lower than that of the 1144 single crystals [2], at around $10^3$-$10^4$ A/cm$^2$ (0 T and 5 K) [11, 10]. Based on these findings, we may need to use a different method for the preparation of 1144 bulks in order to maintain the phase purity and simultaneously improve the grain connections. Recent studies based on the high-pressure synthesis of other FBS families have depicted the improvement of sample quality, crystal size, and the enhancement of the superconducting properties of FBS [16, 17, 18, 19]. These studies motivated us to apply the high-pressure synthesis method to this stoichiometric 1144 materials and investigate the impact of impurity phases on their superconducting properties. As far as our knowledge, no studies have been reported.

In this paper, CaKFe$_4$As$_4$ (1144) bulks are prepared using both the high-pressure and high-temperature synthesis (HP-HTS) approach [20] at different conditions and the conventional synthesis procedure (CSP) at ambient pressure in order to comprehend the high-pressure synthesis and the impurity effect. To conclude our findings, structural, microstructural, transport, and magnetic studies provide a thorough characterization of these samples. Our detailed investigation has depicted that the preparation of high-quality 1144 samples depends on the optimal and each step of the growth conditions to depict high superconducting properties, either by CSP or HP-HTS method. The analysis of several 1144 bulks prepared by CSP and HP-HTS reveals that the superconducting onset transition of these bulks is robust to the presence of 122 impurity phases. Additionally, high-pressure synthesis increased the sample density, critical current density, and critical transition temperature ($T_c$) of this stoichiometric superconductor.

**Experimental details**

Polycrystalline CaKFe$_4$As$_4$ samples were prepared from the initial precursors: Calcium powder (purity 99.95%), arsenic chunks (purity 99.999%), potassium (purity 99.9%), and iron powder (purity 99.99%), as discussed elsewhere [11]. In the first step, CaAs was prepared by using Ca and As heated at 860°C for 30 hours, whereas KAs was prepared by using K and As heated at 650°C for 12 hours. We have also synthesized the Fe$_2$As precursor by heating at 700°C for 12 hours. The precursors were mixed according to the stoichiometric formula of CaKFe$_4$As$_4$. It was then sealed in a Ta-tube under an argon atmosphere through an ARC melter and then in an evacuated quartz tube. Since this 1144 phase is very sensitive to synthesis temperature, as mentioned in the previous study [11], we need to seal the mixed precursors according to CaKFe$_4$As$_4$ composition into a Ta-tube by an ARC melter very carefully, so that the precursor temperature cannot be enhanced to a very high temperature (>50°C) during the tube sealing process. Otherwise, the impurity phases such as CaFe$_2$As$_2$, KFe$_2$As$_2$, can appear with the 1144 phase. To avoid the impurity phases, we generally keep the samples at 955°C for 6 hours, then quench the samples to room temperature by using the ice water. To understand the effect of each step during the synthesis process, CaKFe$_4$As$_4$ bulks were prepared by considering the three synthesis steps: *1)* the best condition, *i.e.*, a quick sealing process of the metal tube through an ARC melter and a quick quenching process, as mentioned in Ref [11] *i.e.* parent sample; *2)* a long time for the metal tube sealing process *i.e.* P1 sample; and *3)* a slow quenching process *i.e.* P2 sample. Also, some samples were prepared at different synthesis temperatures (900°C and 930°C) by CSP to confirm the best conditions *i.e.* P3 and P4 samples. The details about the synthesis conditions, sample code and synthesis process are mentioned in Table 1.

We have also used the high gas pressure and high-temperature synthesis (HP-HTS) method for the preparation of the 1144 phase, as mentioned in Table 1. Our HP-HTS technique is a kind of hot-isostatic pressure (HIP) technique, and it belongs to the gas pressure medium [20]. This approach provides a large sample space (~15 cm length, dia~30–35 mm) by achieving homogenous temperature and pressure stability for a practically long reaction time. There is no possibility of introducing impurities from the pressure medium and we can produce a large amount of samples without being constrained by their size or quantity under the high-pressure conditions. This method has the capability to produce inert gas pressures of up to 1.8 GPa in a cylindrical chamber equipped with a furnace that can achieve temperatures as high as 1700°C [20]. To understand the high-pressure synthesis effect, various CaKFe$_4$As$_4$ were prepared using an *ex-situ* process, i.e., in the first step, the 1144 samples were synthesized by CSP, and in the next step, they were placed into a Ta-tube, either open or sealed, under an inert gas atmosphere and then transferred to the high-pressure chamber. In our previous investigations [17], we prepared numerous samples under various pressure conditions by the HP-HTS process to optimize the high-pressure growth conditions for 11 family, *i.e.*, FeSe$_{0.5}$Te$_{0.5}$ [17, 19]. We found that 500 MPa for one hour was the optimal condition, which was sufficient to produce high-quality FeSe$_{0.5}$Te$_{0.5}$ (11) samples with improved superconducting properties. Therefore, these optimum circumstances were taken into account for the synthesis of 1144 bulks here. We have used the synthesis conditions of 500°C, 500 MPa, and 1-2 hours for the preparation of CaKFe$_4$As$_4$ bulk samples through the HP-HTS process in order to understand the impact of the high-pressure synthesis effect. Since previous studies have suggested that the 1144 samples prepared by CSP are stable up to 600°C [11], we have selected to use a heating temperature of 500°C during the HP-HTS process. Table 1 includes information on the synthesis circumstances as well as a list of samples.

The structure of these prepared bulks was analyzed by X-ray diffraction utilizing a Rigaku SmartLab 3kW diffractometer equipped with filtered *Cu-Kα* radiation (wavelength: 1.5418 Å, power: 30 mA, 40 kV) and a Dtex250 linear detector. To achieve these features, the measurement profile ranging from 5° to 70° was used with a step of 0.01°/min. An additional instrument, an X'Pert PRO, a PANalytical diffractometer with filtered *Cu–Kα* radiation (wavelength: 1.5418 Å, power: 30 mA, 40 kV), and a PIXcel$^{1D}$ position scintillation detector, were also utilised to measure XRD in order to confirm these observations for few samples (Figure S1). Furthermore, the profile analysis was performed with the ICDD PDF4+ 2023 standard diffraction patterns database and Rigaku's PDXL programme. These analyses were used to determine the quantitative values of lattice parameters and impurity phases (%) for different samples. The comprehensive microstructural examination and elemental mapping of these materials are carried out using a Zeiss Ultra Plus field-emission scanning electron microscope equipped with an ultra-fast detector and the EDS microanalysis system by Bruker Model Quantax 400. Utilizing a vibrating sample magnetometer (VSM) attached to a Quantum Design PPMS, the magnetic characteristics of these materials were assessed in the temperature range of 5–45 K and the magnetic field up to 9 T. We used a slow temperature scan in both zero-field-cooled (ZFC) and field-cooled modes to measure the magnetic susceptibility ($\chi$) with a 20 Oe magnetic field. A closed-cycle refrigerator (CCR) was used to measure the temperature variation of resistivity by a four-probe method in a zero magnetic field within the temperature range of 6-300 K. All data were gathered using a very slow warming process.

# Results and discussion

The sample codes with the detailed synthesis conditions of the various CaKFe$_4$As$_4$ compounds are mentioned in Table 1. Figure 1(a) depicts the XRD pattern of all 1144 samples prepared by CSP at ambient pressure. The sample "parent" was prepared very carefully by following all steps as discussed in our previous paper [11, 21]. During the tube sealing process, the sample was carefully cool down through the water surrounding the metal tube, so that the temperature of the precursors inside the metal tube was maintained below 50°C. During the quenching process, we transferred the quartz tube after completing the reaction from the furnace to the ice water very quickly and broke the quartz tube inside the water for a fast cooling. The XRD pattern of this sample did not show any clear peak related to the impurity phase within the machine resolution and confirms the formation of 1144 phase without any impurity phase, as depicted in Figure 1(a). It has a body-centered tetragonal structure with the space group *I4/mmm* [8]. One can note that a very careful analysis of the XRD data of the parent sample can suggest the presence of a very tiny peak of CaFe$_2$As$_2$ phase, estimated to be around ~2-3%. However, confirming this finding is difficult because of its resemblance to the background signal of the XRD measurement. To understand more clearly the effect of the metal tube sealing process or quenching process on the sample growth, we prepared two more 1144 samples at ambient pressure, *i.e.*, P1 and P2. The sample P1 took more time during the Ta-tube sealing process, and it could be possible that this sample faced high temperature during the tube sealing process. One can note that ARC melter can reach a heating temperature of up to 3500°C [22] during the metal tube sealing process. The XRD measurement of this sample shows a huge amount of impurity of CaFe$_2$As$_2$ phase as shown in Figure 1(a) (Table 2). As it is reported [11], if the synthesis is higher than 955°C, commonly most of the impurity phase CaFe$_2$As$_2$ appears during the 1144 synthesis process. Hence, it could be possible to observe a huge amount of CaFe$_2$As$_2$ phase for this sample P1 in the XRD. During the synthesis process of the sample P2, the sample was slowly transferred from the furnace to the ice water, and the quartz tube was not broken inside the water, i.e., it took a slightly longer time for the quenching process of this sample compared to the parent and P1. However, the tube sealing process for this sample P2 was performed in a very short time to avoid heating the Ta-tube, similar to the sample 'parent'. This sample has a main phase of CaFe$_2$As$_2$ and a small amount of FeAs phase, Figure 1(a), which could be due to the effect of a long time for the quenching process from the furnace to room temperature. Interestingly, the samples P2 and P1 both have a large amount of CaFe$_2$As$_2$. To recheck these samples, XRD of these P1 and P2 samples was also performed by a different machine, which is shown in Figure S1 in the supplementary file. To be more clear about the synthesis conditions, we have prepared 1144 at different temperatures of 900°C and 930°C by considering all steps carefully as that for the sample Parent, i.e., P3 and P4. The depicted XRD pattern of these samples have many impurity phases, including the precursors such as $Fe_2As$, $CaFe_2As_2$ and $CaFe_4As_3$, as well as the main phase of $CaKFe_4As_4$, which suggests that these temperatures are not sufficient to complete the reaction for the phase formation of 1144. These results are well in agreement with the previous studies [11, 21]. Furthermore, the samples P3 and P4 suggest that 955°C and 6 h are the best synthesis conditions for the 1144 phase formation. These structural analyses confirm that to prepare a pure phase 1144 sample and all synthesis processes must be performed very carefully, as mentioned for the sample "parent". If the sample synthesis temperature is higher or lower than 955°C during the reaction or after completing the reaction, the impurity phases were observed as depicted for the samples P2, P3, and P4. Even during the sealing of the metal tube, the precursor inside the Ta-tube must be cooled down properly. Table 2 shows the lattice parameters and the amount of impurity $CaFe_2As_2$ phase. The mentioned lattice parameters are almost the same as those of previous reports [8, 11]. One can note that the samples containing impurity phases could have a large error bar in the calculation of lattice parameters.

As a first stage, we have employed our best sample "parent" to comprehend the high-pressure synthesis effect. The sample was then put inside the high-pressure chamber, either sealed or open in a Ta-tube, and HP-HTS was carried out under the synthesis parameters of 500 MPa, 500°C, and 1-2 hours, as was covered in the experimental section. Table 1 lists the sample code and synthesis conditions for these samples. These samples' XRD pattern is displayed in Figure 1(b). The sample HIP_1, prepared in an open Ta-tube, has almost the same XRD pattern as that of the parent, and no impurity phase is detected through XRD measurements. It means the open sample does not degrade the 1144 phase under high-pressure synthesis. Nevertheless, a meticulous examination of the X-ray pattern of HIP_1 yields a comparable outcome to the parent sample, indicating the existence of a tiny peak in the $CaFe_2As_2$ phase (~2%). When the 1144 sample was heated in an open Ta-tube, and in the next step, this sample was sealed in a Ta-tube, i.e., the sample HIP_2, a small amount of $CaFe_2As_2$ and $KFe_2As_2$ phases were observed compared to that of the sample parent and HIP_1. To further confirm the tube sealing effect under pressure, the parent sample was sealed in a Ta-tube by an ARC melter and then placed into the high-pressure chamber of HP-HTS, i.e., the sample HIP_3. This sample was heated at 500°C for 1 hour. Interestingly, the impurity phase of the $CaFe_2As_2$ phase is significantly increased, as depicted in Figure 1(b). At the same time, an extra phase appeared, i.e., $CaFe_4As_3$, as observed for the samples P3 and P4. To be more clear, one more sample was sealed in a Ta-tube placed into the high-pressure chamber, and heated at 500°C, 2 h, and 500 MPa, *i.e.*, HIP_4. The XRD pattern is depicted in Figure 1(b), which display a further increase of the impurity phase. These experiments suggest that the 1144 sample sealed in a Ta-tube under the HP-HTS process is not an effective way to improve the formation of the 1144 phase. The amount of impurity $CaFe_2As_2$ and lattice parameters are mentioned in Table 2. The observed lattice parameters for these samples are almost identical to the previous reports [11, 8]. It proposes that during the growth process of 1144, the phase separation of different phases [23, 24] can occur, if the synthesis conditions are not suitable according to the 1144 phase formation. One can note that 1144 has a very narrow temperature window (~955°C) for a pure phase formation, as reported for the bulk [11] and single crystal [14].

The elemental mapping of different samples is carried out in order to understand the distribution of the constituent elements inside these bulk samples. The elemental mappings of parent, P2, HIP_1, and HIP_3, are shown in Figure 2. Figure 2(i) illustrates the nearly homogeneous distribution of all the elements Fe, Ca, K, and As in the parent sample, indicating the well-formed 1144 phase and corroborated by the XRD data. However, very few and small areas appear to be rich in Ca, Fe and As (Figure 2(i)), suggesting the possible presence of a tiny amount of $CaFe_2As_2$ phase, as proposed from XRD analysis. Sample P2 displays the non-homogeneous distribution of the constituent elements (Figure 2(ii)), and a few locations exhibit Ca, Fe, and As-rich regions, indicating the phase development of $CaFe_2As_2$, FeAs, or $Fe_2As$. For the samples P3 and P4, which are displayed in Figure S2 (supplementary file), a non-homogeneous distribution of the elemental mapping is also more strongly detected (Figure S2). It's interesting to note that the elemental mapping of HIP_1 has an almost homogeneous distribution of all elements, but small K-enriched regions can be clearly observed in the upper corner of Figure 2(iii). On average, the elemental mapping of HIP_1 is comparable to the parent compound, as shown in Figure 2(i), and proposes the existence of a very minute amount of $CaFe_2As_2$ or $KFe_2As_2$ phase that could not be detected in XRD analysis. In the case of the sample sealed into a Ta-tube, *i.e.,* the sample HIP_3, the elemental mapping is shown in Figure 2(iv). Numerous locations were found to be rich in Fe and As, while a small number of locations were found to be Ca-rich, indicating the creation of the $CaFe_2As_2$ or $CaFe_4As_3$ and FeAs phases. According to these studies, a long heating period under pressure or the sample sealed into a Ta-tube is not suitable for the 1144 phase, which generates the inhomogeneity of the constituent elements, i.e., the formation of the impurity phases. This analysis is followed by the samples P3, P4, and HIP_4, which are displayed in Figure S2 (supplementary file) and illustrate the enhancement of the non-homogeneous distribution of the component elements. In contrast, for the pure phase formation of 1144, high-pressure synthesis works almost well for the sample placed in an open Ta-tube inside a high-pressure chamber. These mapping analyses corroborate the XRD mentioned previously.

For the microstructural analysis, we have collected back-scattered images (BSE) for various samples. Figure 3(a)-(i) shows the BSE images of the samples: Parent, HIP_1, and HIP_3 ranging from high to low magnifications. The corresponding secondary electron (SE) images for these samples are depicted in Figure S3 as the supplementary data. Using emery paper of several grades and without the use of liquid, these samples were meticulously polished. Light gray and black contrasts were observed corresponding to $CaFe_4As_4$ (1144) and pores, respectively. The sample parent has a fairly homogeneous microstructure, where mainly grey and black contrasts are observed. Many pores do exist, and some well-connected grains have appeared. Based on Figures 3(d)–(e), it appears that sample HIP_1 is relatively compact, with a large number of well-connected grains and a reduced number of pores. Parent and HIP_1 do not exhibit any impurity phases. The sample that was heated under pressure for an extended period, i.e., HIP_2, has reduced grain connections and increased pore size, which is displayed in Figure S4 (Supplementary file). Figure 3(g) illustrates the white contrast on the surface caused by the $CaFe_2As_2$ phase, which may be the result of an unstable phase of $CaFe_2As_2$. Other samples (P3, P4) containing this $CaFe_2As_2$ or $CaFe_4As_3$ impurity phase also exhibit similar whitish contrast. As the $CaFe_2As_2$ phase is air-sensitive, it should be noted that the samples were very briefly exposed to air while being fixed on a SEM holder. A recent research have demonstrated the oxidation of 1144 bulks [25] . In comparison to the parent and HIP_1, it appears that HIP_3 has fewer grain connections and larger pores. As mentioned with XRD and mapping, it might be because of the impurity phase development. Similar to the HIP_3 sample, white contrast is increased and observed in the larger area, and the grain connections are reduced for HIP_4 and other samples P2 and P4 (Figure S4). Figure S5 displays the SE images of the following sample: P2, P4, HIP_2 and HIP_4. Taking into account the theoretical density of 5.22 g/cm$^3$ for the 1144 phase [8, 11], the following calculated densities are observed: 66%, 35%, 77%, 51%, 45%, and 42% for the parent, P2, HIP_1, and HIP_2, HIP_3, and HIP_4, respectively. These calculated densities closely match the results of the observed microstructural analysis.

The resistivity behavior of all 1144 samples prepared by CSP is illustrated in Figure 4(a) from the room-temperature to low-temperature data. In the normal condition, all of these samples exhibit linear resistivity behaviour, which is consistent with earlier reports [9, 13, 11]. In comparison to other samples, the parent sample exhibits the lowest resistivity, indicating a clean phase of 1144 that is consistent with XRD data and behaves similarly to the previous reports [8, 10, 9, 11]. Remarkably, the resistivity of samples P1 and P2 is marginally higher than that of the parent sample across the complete temperature range, suggesting the presence of impurity phases. More intriguingly, sample P2 exhibits a very slight increase in resistivity and behaves similarly to the pure sample, while having a major phase of the impurity $CaFe_2As_2$. This sample's normal state resistivity has a kink at approximately 170 K, which is caused by the structural and magnetic transition of $CaFe_2As_2$ phase [14]. The samples P3 and P4 have many other impurity phases with the $CaFe_2As_2$ phase and have shown a very high resistivity value compared to that of the parent sample. It could be due to the presence of many impurity phases and suggests that the chemical reaction is not completed for the phase formation of the 1144, as well agreed with the previous report [11]. One can note that the samples P1 and P2 have a large amount of $CaFe_2As_2$ phase, but it does not affect too much the resistivity behaviour or the value of the resistivity compared to the parent compound. One can note that a small kink around 170 K is also observed for the parent sample, which is more clearly visible for P1, P2, P3, and P4, where the $CaFe_2As_2$ phase was clearly observed by XRD measurements, as discussed above. To check more carefully, we have measured the resistivity of different pieces of the parent sample belonging to one batch sample, as depicted in the supplementary Figure S6. Interestingly, we have observed a very tiny peak or almost no visible kink around 170 K for these pieces (Figure S6), which supports the presence of a very small amount of $CaFe_2As_2$ phase from the elemental mapping, but it could not detectable from the XRD data for the parent sample. It suggests that the presence of $CaFe_2As_2$ has little effect on the resistivity behaviour of the 1144 phase, even though it was presented as a main phase in the sample P2.

The low-temperature behavior of these samples is shown in Figure 4(b). The parent compound shows an onset transition temperature of 33.5 K with a transition width of 1 K. The samples P1 and P2 have almost the same onset transition as that of the parent compound, even though these samples have a large amount of $CaFe_2As_2$ as an impurity phase. However, the sample P1 has a transition width of ~1 K as it has around 46% impurity of 122. The transition temperature is broader (~4 K) for the sample P2, where the 122 phase is observed as a main phase (77.56 %). It has a slightly broader transition width than the parent sample. The samples P3 and P4 have many impurity phases ($CaFe_2As_2$, $CaFe_4As_3$, FeAs), but interestingly, the onset superconducting transition is again almost the same as that of the parent sample. The only noticeable point is a huge broadening of transition width and a higher resistivity value compared to the parent sample. Again, these results suggest that there is a phase separation of 122 and 1144 phases during the synthesis process of these samples. However, the presence of the impurity phases such as the 122 phase does not affect too much on the superconducting properties of 1144. These results are completely different from the behavior of other iron-based superconductors [23, 24], where the presence of an impurity phase reduces the onset superconducting transition. However, here, the presence of the impurity 122 phase as a main phase in the sample P2 has the same onset transition temperature as that observed for a pure 1144 sample.

Figure 5(a) shows the temperature dependence of the resistivity behavior of the 1144 bulks prepared by HP-HTS with the best 1144 (parent) obtained by CSP. All the samples prepared by HP-HTS have similar behavior to the parent sample. The normal state resistivity value of the sample HIP_1 is slightly lower than the parent sample, which could be possible due to the slightly improved sample density. The resistivity is increased for the sample HIP_2 in the whole temperature range due to the presence of the impurity, which reduces the sample density. A slight kink in the resistivity around 170 K is observed due to the $CaFe_2As_2$ phase, as reported for the 1144 single crystal [14]. The sample sealed into a Ta-tube and heated for 1 hour, i.e., HIP_3, has increased the resistivity value due to the impurity phases $CaFe_2As_2$ and FeAs, as depicted in the inset of Figure 5(a). This trend is also followed by the sample HIP_4 heated for a longer time (~2 hours) under high pressure, and it has shown a huge enhancement of the resistivity due to the presence of many impurity phases, as discussed with XRD analysis. A small kink for HIP_1 is also observed around 170 K, suggesting the presence of a tiny amount of $CaFe_2As_2$ phase similar to the parent sample and supporting its elemental mapping analysis (Figure 2(iii)).

The low-temperature behavior is shown in Figure 5(b). The HIP_1 sample has enhanced the $T_c^{onset}$ by 2 K and reached up to 35.5 K with a transition width of 1 K. These values are the same as what was reported for 1144 single crystals and suggests a good quality sample with good grain connectivity, which is well agreed with the microstructural analysis. Other samples, such as HIP_2, HIP_3, and HIP_4, have almost the same onset $T_c$ as that of the parent sample, but the transition width ($\Delta T$) is increased with the amount of the impurity phase. These analyses again suggest the presence of impurity phases does not affect the onset value of the superconducting transition of the 1144 phase, as similar to the samples prepared by CSP and supports the phase separation between the 122 and 1144 phase during the synthesis process.

To confirm the Meissner effect of these samples, the magnetic susceptibility was measured in ZFC and FC mode for the different samples as depicted in Figure 6(a). To make a comparative study, we have shown the normalized magnetization for various samples. The parent and P2 samples have shown the onset transition of 33.3 K which is almost the same as that of the resistivity measurements. The enhancement of transition is also clearly observed for HIP_1 with a $T_c$ value of 34.5 K. The sample HIP_3 has an onset $T_c$ value of 33 K, as similar to the parent and P2 sample, but has a broader transition. The two-step transition behaviour for these samples could be due to a feature of granular magnetic response that is well-known for iron- based superconductors [26, 7, 5]. These measurements also confirm that the presence of the impurity phase does not affect the onset transition value of a pure 1144 sample but enhances the broadening of the transition. This is well in agreement with the above observations from the transport measurements. Also, our studies confirm that HP-HTS process work well to improve the superconducting properties under suitable pressure conditions and it is possible to prepare a large amount of the bulk sample in one batch through this technique, which can be useful for the fabrication of superconducting wires and tapes [15].

For the practical application of a superconductor, the critical current density plays a crucial role. The magnetic hysteresis loop at 5 K is measured for three samples: parent, P2, and HIP_1 in the presence of the magnetic field of 9 T, as shown in the inset of Figure 6(b). The rectangular-shaped bulks were used for the magnetic measurements. The observed hysteresis loops are similar to the previous reports [10, 11]. The hysteresis loop width ($\Delta m$) for these bulks using their hysteresis loops can be calculated by variation of magnetization when sweeping the magnetic field up and down. Bean model is used to calculate the critical current density $J_c = 20\Delta m/Va(1-a/3b)$ [27] by using the hysteresis loop ($\Delta m$), where $a$ and $b$ are the short and long edges of the sample ($a < b$), and $V$ is the volume of the sample. The calculated critical current density with the variation of magnetic field for these samples is shown in Figure 6(b) at 5 K. HIP_2 samples were very small and thin, so it was hard to make a rectangular shape. Due to this, we have not included the $J_c$ calculation in this figure. Based on the structural and microstructural analysis of this sample, a low value of $J_c$ can be expected due to the presence of the impurity phases and low sample density. Figure 6(b) depict a small variation of $J_c$ with the magnetic field up to 9 T and the obtained $J_c$ value for the parent sample is ~ 7.5 × $10^3$ A/cm$^2$ at 0 T and 5 K, which is slightly lower than the previous report of 1144 bulk. Since this parent sample has the sample density of 66% which is much lower than that (~99%) of the previous report [11], it might be a reason for the lower $J_c$ value for the parent sample. The lower sample density and the presence of impurity reduced $J_c$ value as observed for the sample P2. The 1144 bulks prepared by high pressure synthesis process i.e. HIP_1, has an enhanced $J_c$ by one order of magnitude (6 × $10^4$ A/cm$^2$ at 0 T) than that of the parent sample and almost same as that the previous reports [9, 11]. It could be due to the improvement of sample density of CaKFe$_4$As$_4$ by HP-HTS. Furthermore, the improvement in material density, grain connections, and suitable pinning centres as observed for FBS [26, 7, 28] and MgB$_2$ [29] may be the cause of this increase in $J_c$ by HP-HTS, indicating that high-pressure synthesis is an effective method for enhancing intergrain connections and improving pinning properties.

To summarize our findings, we have plotted the amount of impurity phase (CaFe$_2$As$_2$), the onset transition temperature ($T_c^{onset}$), room temperature resistivity ($\rho_{300K}$), residual resistivity ratio (RRR = $\rho_{300K}$ / $\rho_{40K}$) and the transition width ($\Delta T = T_c^{onset} - T_c^{offset}$) for 1144 bulk prepared by CSP and HP-HTS, as depicted in Figures 7 and 8. Supplementary Figures S7 and S8 display an alternative version of these graphs. Different samples (Parent, P1, and P2) have different amounts of CaFe$_2$As$_2$ phase during the preparation of CaKFe$_4$As$_4$ composition under the same synthesis conditions, as depicted in Figure 7(a). The P3 and P4 have a lower amount of CaFe$_2$As$_2$ phase than that of P1 and P2, but at the same time, other impurity phases (FeAs, CaFe$_4$As$_3$) are observed (Figure 1). The onset $T_c$ value is the same for the Parent, P1, and P2 samples, which suggests the impurity phase does not affect the onset value of the superconducting transition, whereas the samples P3 and P4 have a lower $T_c^{onset}$ value (Figure 7(b)). The variation of room temperature resistivity ($\rho_{300K}$) for these samples is shown in Figure 7(c) and suggests that the presence of CaFe$_2$As$_2$ does not affect too much the normal state resistivity value and its behaviour, i.e., for the parent, P1 and P2. However, this $\rho_{300K}$ is increased for the samples P3 and P4, where many impurity phases do exist. *RRR* value generally gives information about the homogeneity of the sample and well grain connectivity. The sample parent and P1 have almost the same *RRR* value, as in Figure 7(d), even the sample P1 has 46% CaFe$_2$As$_2$ phase. The sample P2 has the main phase of CaFe$_2$As$_2$ and has a reduced *RRR* compared to all other samples. As the phase formation of 1144 is increased from 900 to 930°C, RRR is slightly reduced, as shown in Figure 7(d), but smaller than that of the parent sample. The width of the transition is shown in Figure 7(e) for various samples and is almost the same for the parent and sample P1. However, it is enhanced for the samples P2, P3, and P4 with the enhanced the impurity phases. These analyses suggest that the presence of the $CaFe_2As_2$ phase does not affect the onset superconducting transition but reduces the sample quality and grain connections, although not so effectively. It could be due to the phase separation of different phases, when the synthesis conditions are not very suitable for the formation of the 1144 phase.

Figure 8(a)-(f) depicts the variation of various parameters for different samples prepared by HP-HTS with respect to the parent sample prepared by CSP. Almost no impurity phase was observed for Parent and HIP_1. The $CaFe_2As_2$ phase is increased for a long-heated sample under pressure, i.e., for HIP_2 and HIP_4 samples, however, it is slightly lower than the sample HIP_3. The onset $T_c$ is reached up to 35.3 K for HIP_1, as shown in Figure 8(b), which is similar to the reported single crystal [13, 2] and 2 K higher than that of our parent sample. The sample sealed in a Ta-tube or heated for a long time under pressure reduces the $T_c$ value, which is slightly lower than that of the parent bulks. It is again suggested that the onset $T_c$ is not very affected by the amount of $CaFe_2As_2$ phase. The room-temperature resistivity value is slightly different for HIP_1 and HIP_2, but it increases for HIP_3 and HIP_4, which could be due to the presence of many impurity phases. The RRR value is depicted in Figure 8(d). HIP_1 has a slightly improved RRR value due to improved sample density and good grain connectivity, whereas it reduces for other samples compared to the parent and HIP_1 samples. One can note that the RRR value generally decreases for the samples containing the impurity phases. The enhancement of the impurity phases increases the broadening of the onset transition, Figure 8(e), and this behaviour is the same as observed for the samples prepared by CSP. The transition width is almost the same for the open and closed samples, i.e. HIP_1 and HIP_3, as shown in Figure 8(e). It is increased for the long-heated samples, i.e., HIP_2 and HIP_4. The calculated $J_c$ is shown in Figure 8(f) for the parent and HIP_1. The sample prepared by HP-HTS has shown an enhancement of the $J_c$ value compared to the parent sample prepared by CSP. It suggests that HP-HTS improved the pinning centers and the sample density. Hence, the 1144 sample heated in an open metal tube is effective for improving of its superconducting properties, whereas the sample sealed in a Ta-tube or heated for a long time under the pressure exhibits the deterioration of the superconducting properties of the 1144 phase. Furthermore, this study also suggests that HP-HTS can be a unique technique for further improving the superconducting properties of 1144 family; however, we need more studies in this direction, such as the optimization of the synthesis of 1144 by HP-HTS by considering the various parameters [17, 21].

## Conclusion

We prepared CaKFe$_4$As$_4$ bulks using CSP and HP-HTS methods by considering various synthesis conditions. Our study suggested that the preparation of the 1144 phase is very sensitive with respect to each step of the synthesis process. The metal tube sealing process by ARC melter before the reaction and a quenching process after completing the reaction also play an important role in the pure phase formation of CaKFe$_4$As$_4$. The common impurity phases, especially CaFe$_2$As$_2$ or KFe$_2$As$_2$ (122), did not affect the onset transition temperature ($T_c^{onset}$) of 1144 and are very less effective on the broadening of the superconducting transition ($\varDelta T$). Our findings revealed that when synthesis conditions are unfavourable during the preparation of the 1144 phase, the phase separation effect could be a reason for the phase 122's formation. However, the formation of other impurity phases, such as FeAs or CaFe$_4$As$_3$ effectively reduced the onset transition temperature and enhanced the broadening of this transition. The results of high-pressure synthesis demonstrated that CaKFe$_4$As$_4$ sample placed in an open metal tube exhibited an increase of the transition temperature by 2 K with the sharpening of the transition, which is similar to that of the reported 1144 single crystal. The critical current density of this sample is also enhanced by one order of magnitude compared to the parent sample prepared by CSP. However, CaKFe$_4$As$_4$ bulks sealed in Ta-tube or heated for a long time under high pressure resulted in an increase in the impurity phases and a decrease in their superconducting properties. Superconducting properties and sample quality can be enhanced by preparing CaKFe$_4$As$_4$ bulks in an open Ta-tube for one hour at 500 MPa. Our results confirm that the high-pressure synthesis of 1144 works well, and more study is needed in this direction for the further enhancement of the superconducting properties of the 1144 family.

**Acknowledgments:**

The work was funded by SONATA-BIS 11 project (Registration number: 2021/42/E/ST5/00262) funded by National Science Centre (NCN), Poland. SJS acknowledges financial support from National Science Centre (NCN), Poland through research Project number: 2021/42/E/ST5/00262.

**Table 1:** Details of synthesis conditions, synthesis process and the sample codes for the prepared CaKFe$_4$As$_4$ bulks by conventional synthesis process at ambient pressure (CSP), and high gas pressure and high temperature synthesis (HP-HTS) process, as detailed discussed in the experimental part. Given that HP-HTS is a type of hot-isostatic pressure (HIP) technology, we have used the "HIP" instead of HP-HTS for the sample code as a short name.

| Sample Synthesis conditions | Sample code | Process |
|---|---|---|
| *First step:* heated at 955 °C, 6 h, 0 MPa ↓ *Second step:* heated at 955 °C, 2 h, 0 MPa (Proper sealing and quenching process) | Parent | *In-situ* |
| *First step:* heated at 955 °C, 6 h, 0 MPa ↓ *Second step:* heated at 955 °C, 2 h, 0 MPa (Long time for Ta-tube sealing process) | P1 | *In-situ* |
| *First step:* heated at 955 °C, 6 h, 0 MPa ↓ *Second step:* heated at 955 °C, 2 h, 0 MPa (A slow quenching process) | P2 | *In-situ* |
| *First step:* heated at 900 °C, 6 h, 0 MPa ↓ *Second step:* heated at 900 °C, 2 h, 0 MPa | P3 | *In-situ* |
| *First step:* heated at 930 °C, 6 h, 0 MPa ↓ *Second step:* heated at 930 °C, 2 h, 0 MPa | P4 | *In-situ* |
| 500 °C, 1 h, 500 MPa (Open Ta tube) | HIP_1 | *Ex-situ* |
| *First step*: 500 °C, 1 h, 500 MPa (Open Ta tube) ↓ *Second step*: 500 °C, 1 h, 500 MPa (Close Ta tube) | HIP_2 | *Ex-situ* |
| 500 °C, 1 h, 500 MPa (Close Ta tube) | HIP_3 | *Ex-situ* |
| 500 °C, 2 h, 500 MPa (Close Ta tube) | HIP_4 | *Ex-situ* |

**Table 2:**

A list of the calculated lattice parameters '*a*' and '*c*', and the amount of impurity phase $CaFe_2As_2$ (%) for the prepared $CaKFe_4As_4$ bulks is provided. Rigaku's PDXL software and the ICDD PDF4+ 2023 standard diffraction patterns database have been used for the quantitative analysis of the impurity $CaFe_2As_2$ phase (%) and the lattice parameters through the refinement of the measured XRD data.

| Sample code | Lattice '*a*' (Å) | Lattice '*c*' (Å) | $CaFe_2As_2$ (%) |
|---|---|---|---|
| **Parent** | 3.872(3) | 12.851(2) | ~2 |
| **P1** | 3.872(5) | 12.851(2) | 46.6 |
| **P2** | 3.869(4) | 12.815(2) | 77.56 |
| **P3** | 3.869(1) | 12.862(3) | 30.7 |
| **P4** | 3.869(2) | 12.841(1) | 26.5 |
| **HIP_1** | 3.863(5) | 12.852(3) | ~2 |
| **HIP_2** | 3.852(1) | 12.863(2) | 5.04 |
| **HIP_3** | 3.868(3) | 12.852(1) | 20.3 |
| **HIP_4** | 3.868(4) | 12.838(1) | 27.7 |

**Figure 1:** Powder X-ray diffraction (XRD) patterns for CaKFe$_4$As$_4$ bulks prepared by **(a)** CSP and **(b)** HP-HTS process under different growth conditions.

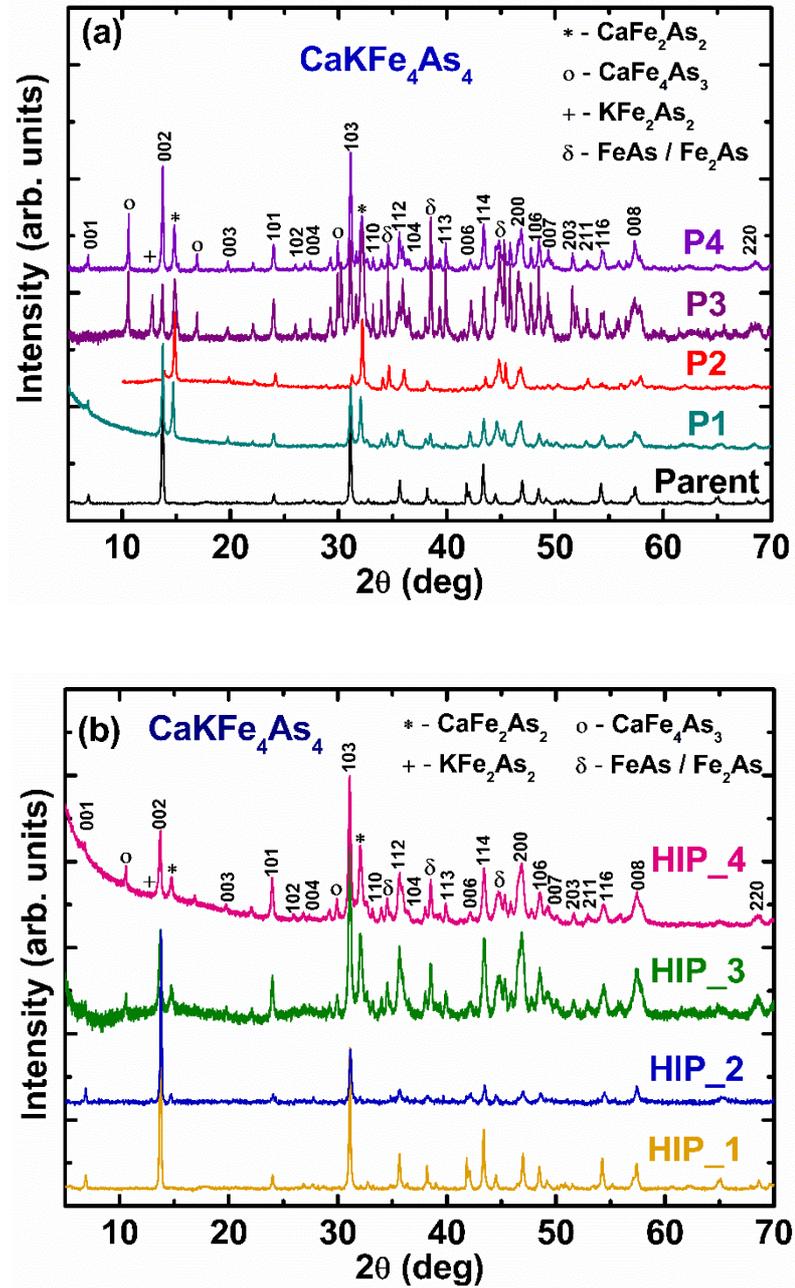

**Figure 2:** Elemental mapping of the constituent elements for various CaKFe$_4$As$_4$ samples: **(i)** parent **(ii)** P2 **(iii)** HIP_1 **(iv)** HIP_3.

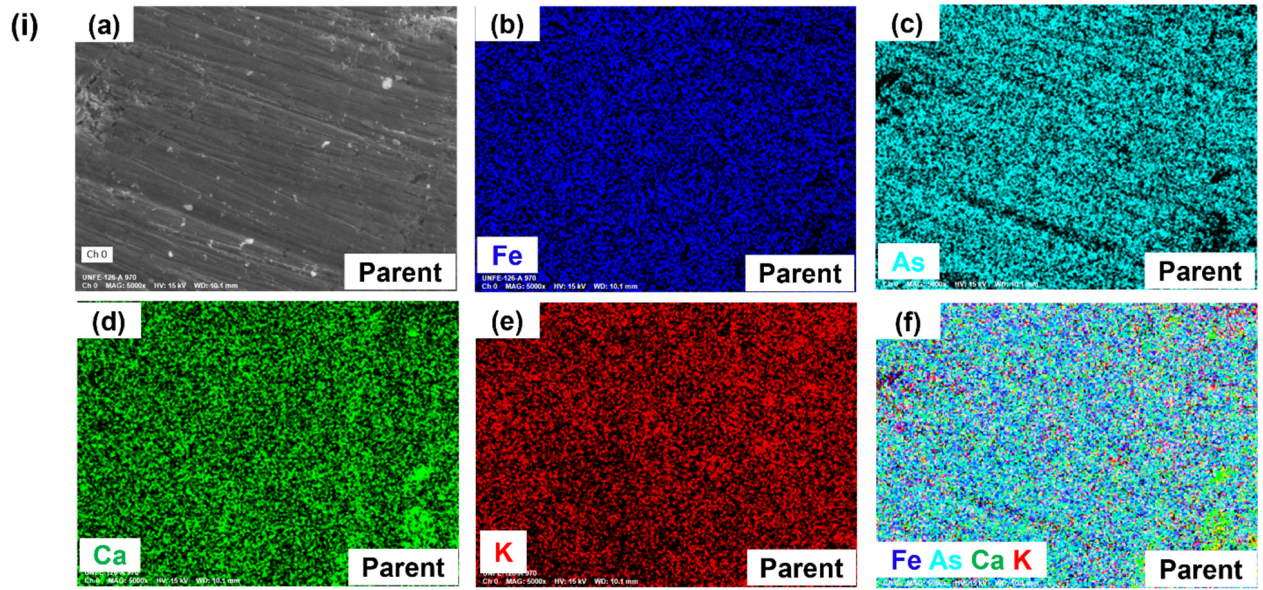

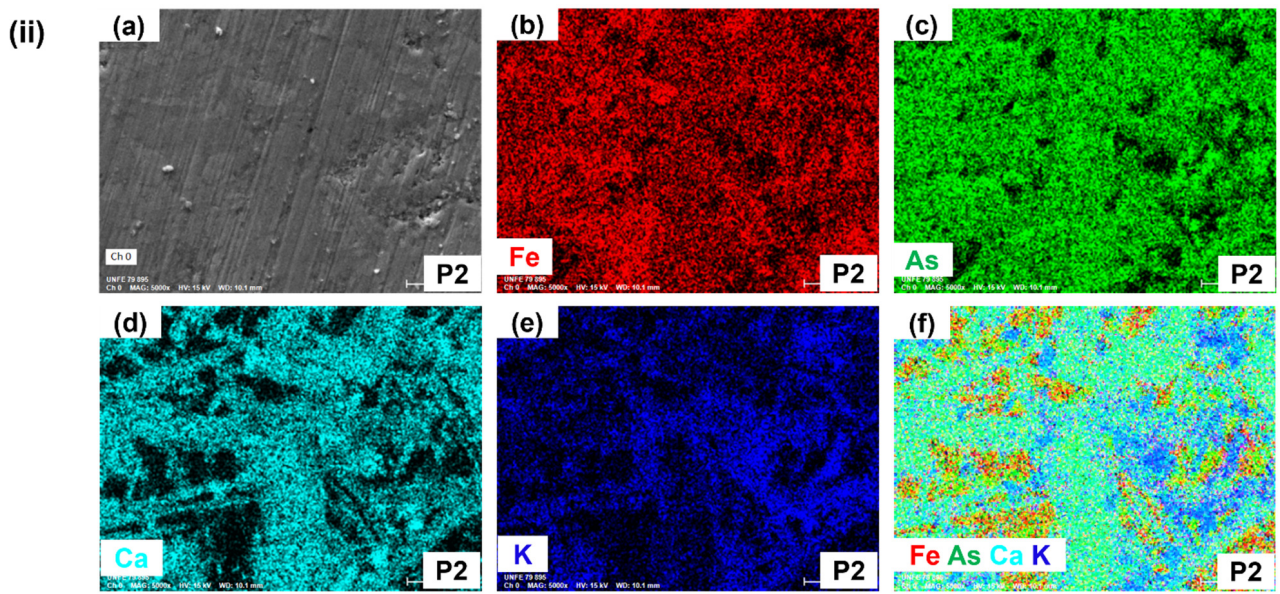

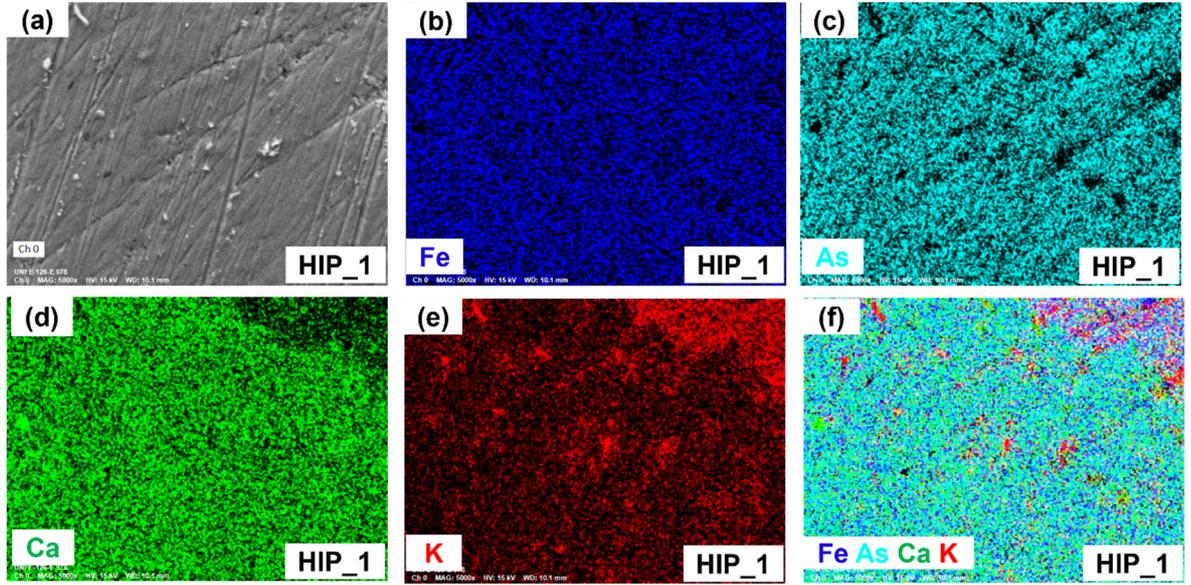

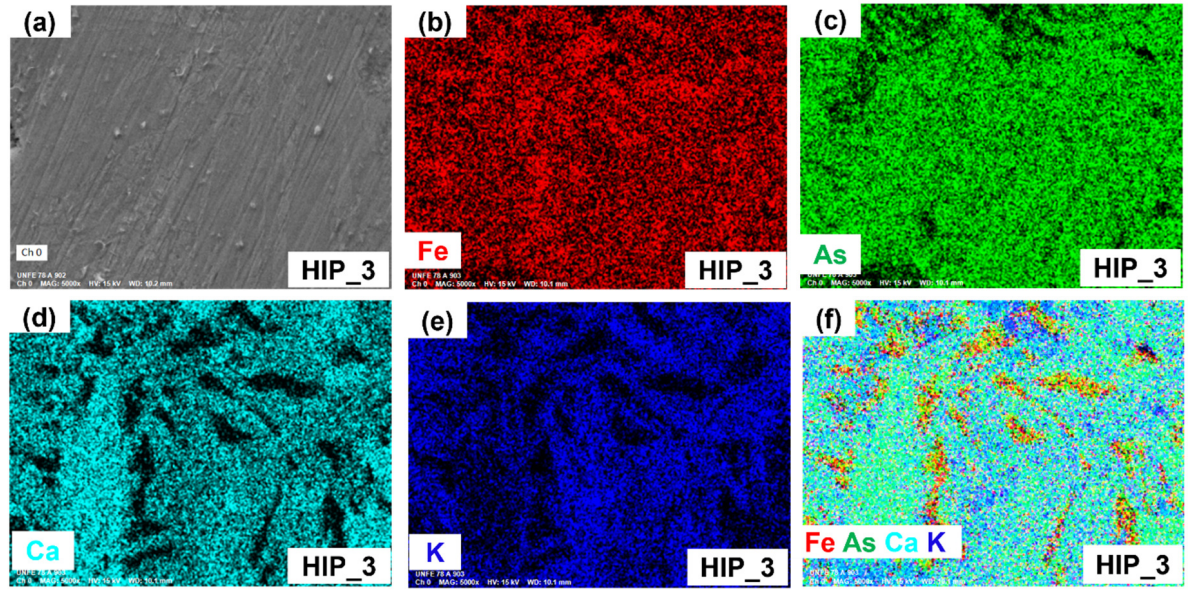

**Figure 3:** Backscattered electrons (BSE: AsB) images from high to low magnifications for CaKFe$_4$As$_4$ sample: **(a)-(c)** parent, **(d)-(f)** HIP_1 and **(g)-(i)** HIP_3.

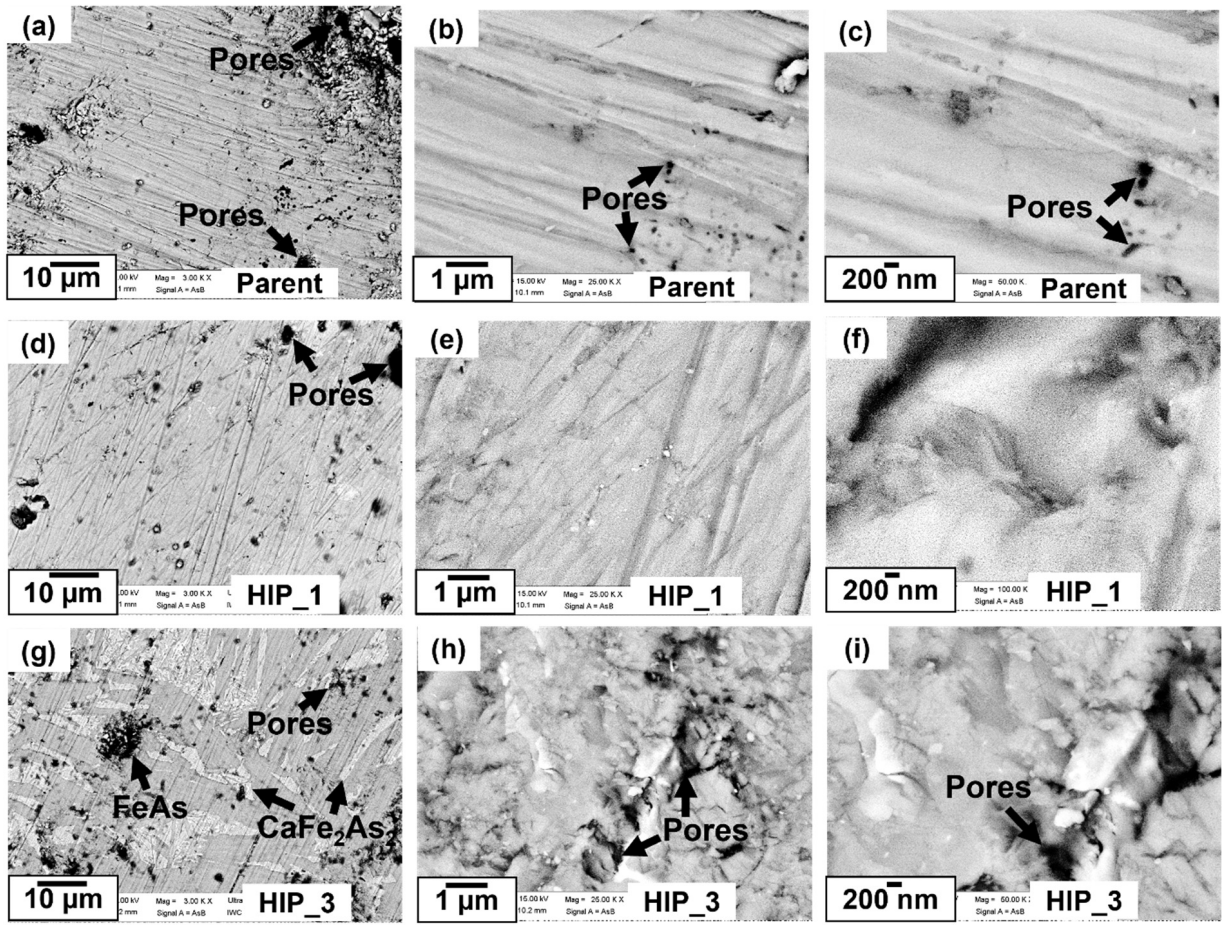

**Figure 4:** (a) The temperature variation of resistivity ($\rho$) up to room temperature for various CaKFe$_4$As$_4$ bulks prepared by CSP. (b) The low-temperature variation of resistivity up to 36 K for these samples.

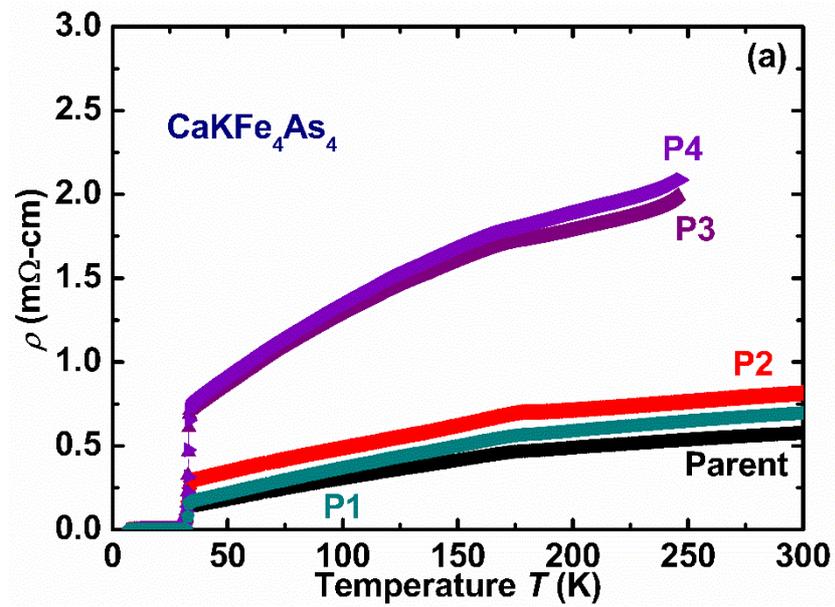

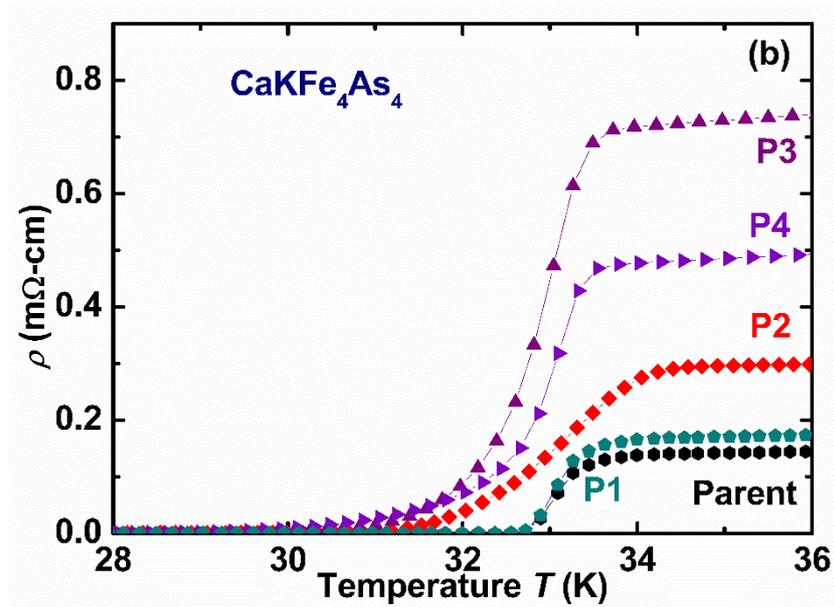

**Figure 5:** (a) The variation of resistivity (ρ) with temperature up to room temperature for various 1144 samples prepared by HP-HTS. The inset figure shows the temperature dependence of the resistivity for HIP_3 and HIP_4. (b) The low-temperature variation of resistivity for these samples up to 40 K.

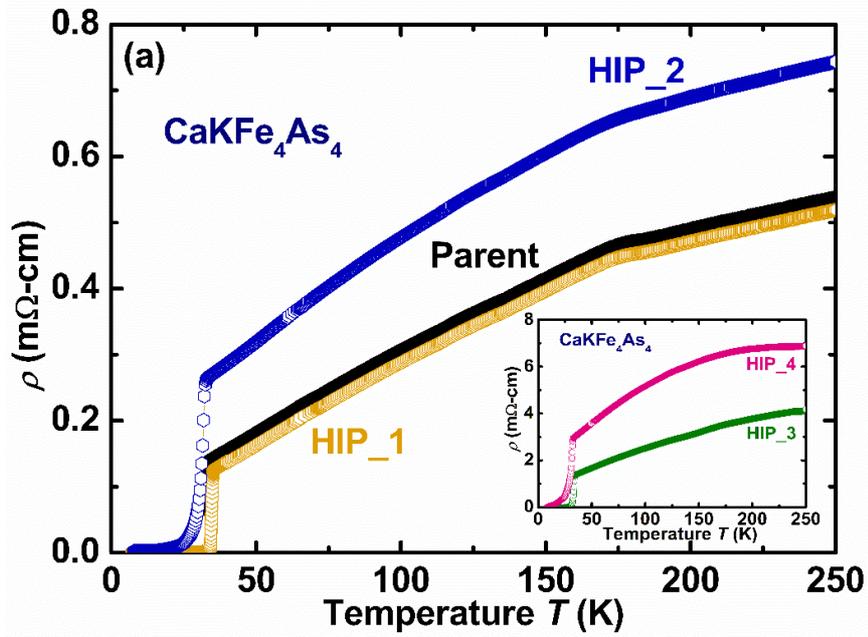

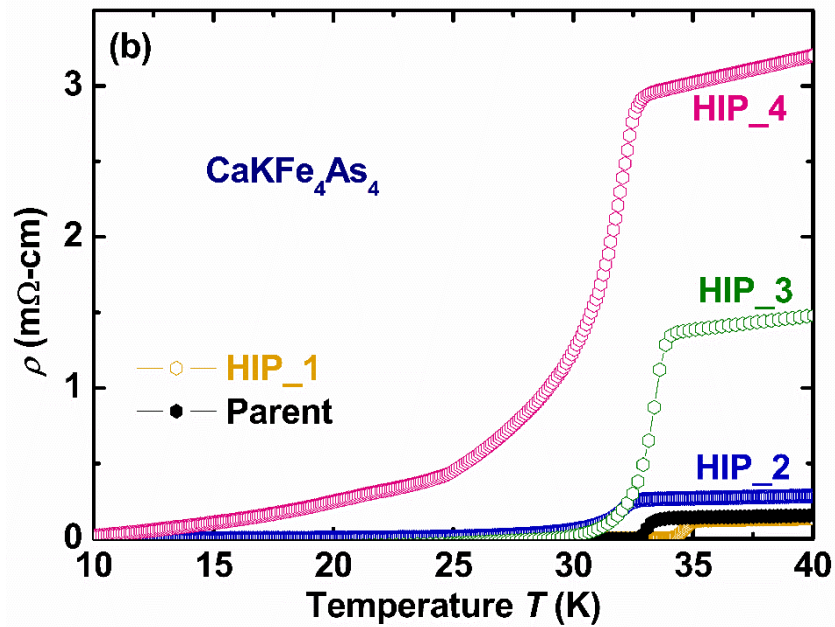

**Figure 6:** **(a)** The temperature dependence of the normalized magnetic susceptibility ($4\pi M/H$) in ZFC and FC mode for various CaKFe$_4$As$_4$ bulks: Parent, P2, HIP_1 and HIP_3 in the presence of 20 Oe magnetic field. **(b)** The magnetic field variation of critical current density ($J_c$) up to the magnetic field of 9 T at 5 K for parent, P2 and HIP_1 samples. The inset figure shows *M-H* loops at 5 K for CaKFe$_4$As$_4$ samples: Parent, P2 and HIP_1.

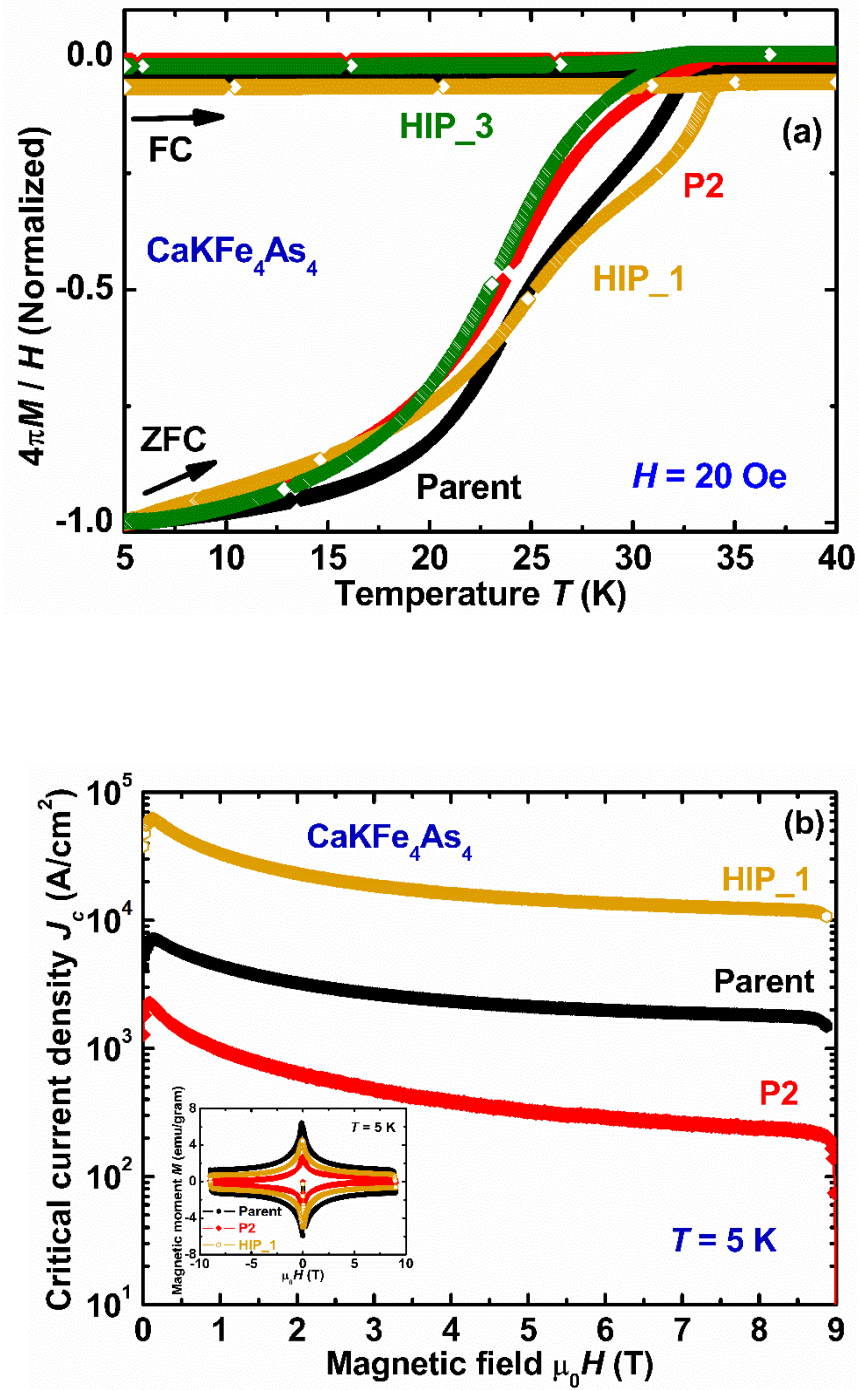

**Figure 7:** The variation of **(a)** an amount (%) of the impurity $CaFe_2As_2$ phase, **(b)** the onset transition temperature $T_c^{onset}$, **(c)** the room temperature resistivity ($\rho_{300K}$), **(d)** Residual resistivity ration ($RRR = \rho_{300\,K} / \rho_{40\,K}$) and **(e)** transition width ($\Delta T$) with respect to various $CaKFe_4As_4$ bulks prepared by CSP.

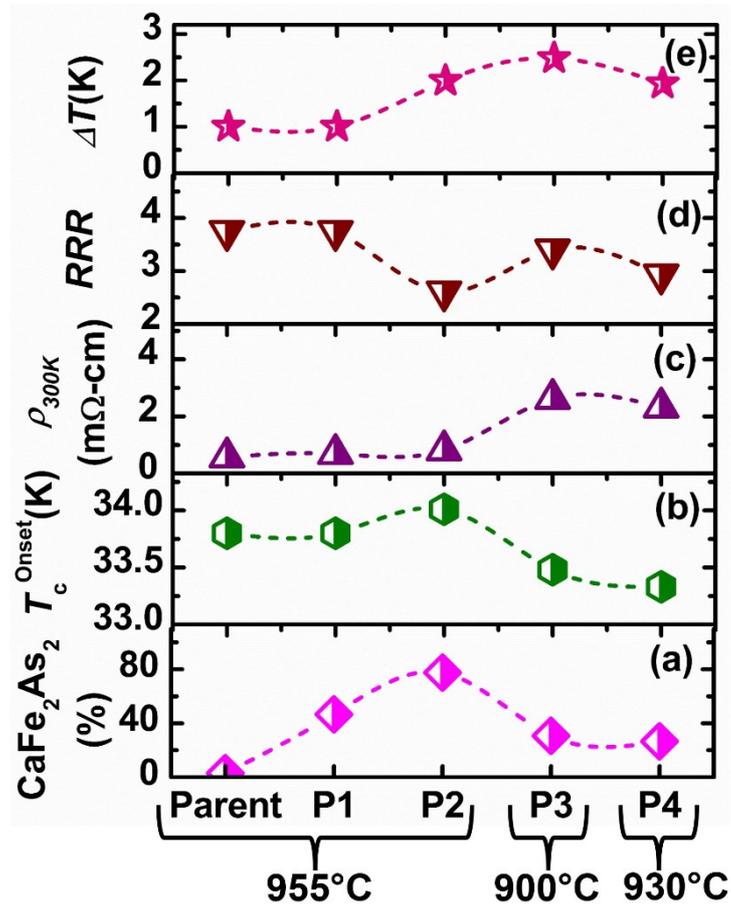

**Figure 8:** The variation of **(a)** an amount of the impurity $CaFe_2As_2$ phase, **(b)** the onset transition temperature $T_c^{onset}$, **(c)** the room temperature resistivity ($\rho_{300K}$), **(d)** Residual resistivity ration ($RRR = \rho_{300\,K} / \rho_{40\,K}$), **(e)** transition width ($\Delta T$) and **(f)** the critical current density ($J_c$) at 5 K and self-field with respect to $CaKFe_4As_4$ bulks prepared by HP-HTS method with the "parent" sample prepared by CSP method.

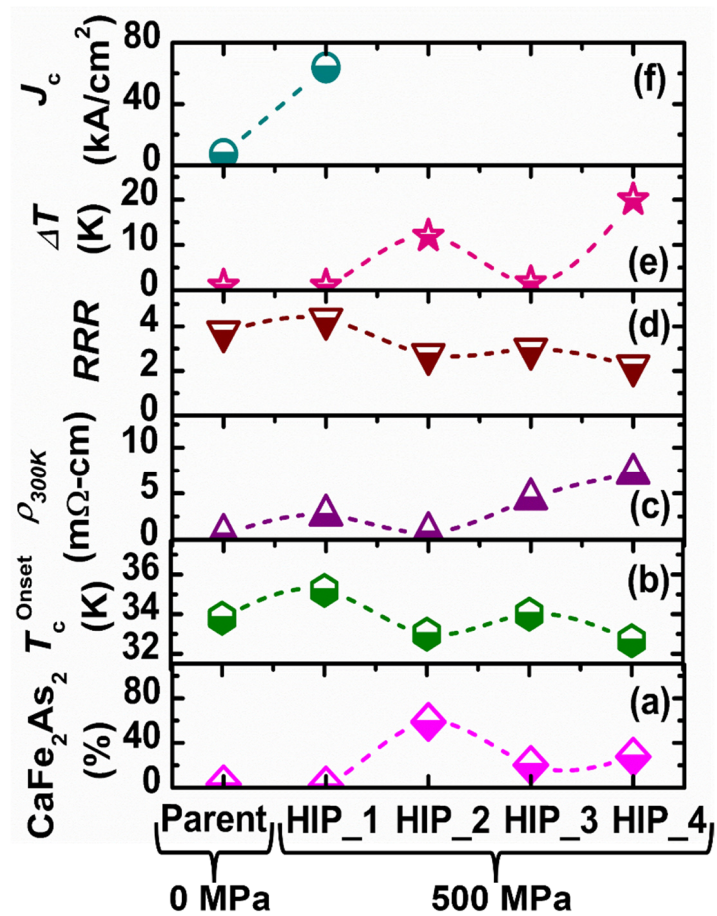

# Effect of impurity phase and high-pressure synthesis on the superconducting properties of CaKFe$_4$As$_4$

Manasa Manasa[1], Mohammad Azam[1], Tatiana Zajarniuk[2], Ryszard Diduszko[3], Tomasz Cetner[1], Andrzej Morawski[1], Andrzej Wiśniewski[2], Shiv J. Singh[1*]

[1]*Institute of High Pressure Physics (IHPP), Polish Academy of Sciences, Sokołowska 29/37, 01-142 Warsaw, Poland*

[2]*Institute of Physics, Polish Academy of Sciences, aleja Lotników 32/46, 02-668 Warsaw, Poland*

[3]*Łukasiewicz Research Network Institute of Microelectronics and Photonics, Aleja Lotników 32/46, 02-668 Warsaw, Poland*

*Corresponding author (Email: sjs@unipress.waw.pl )

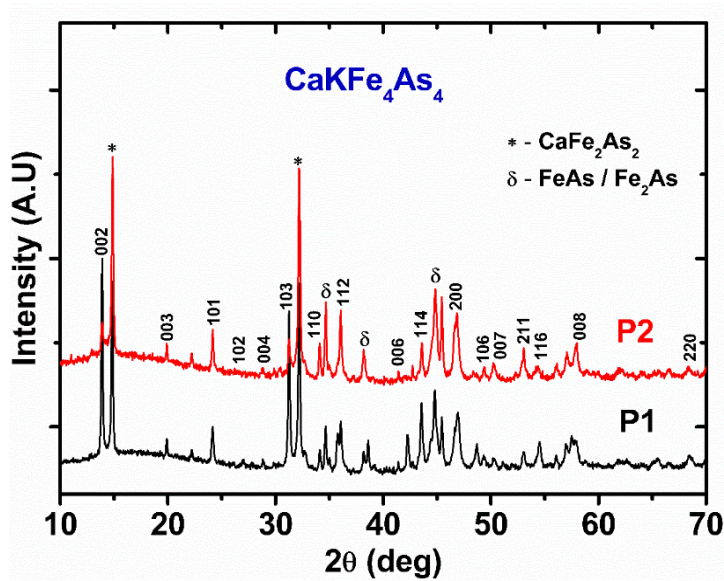

**Figure S1:** X-ray diffraction (XRD) patterns for P1 and P2 bulks prepared by conventional synthesis process at ambient pressure (CSP) using X'Pert PRO and PANalytical diffractometer.

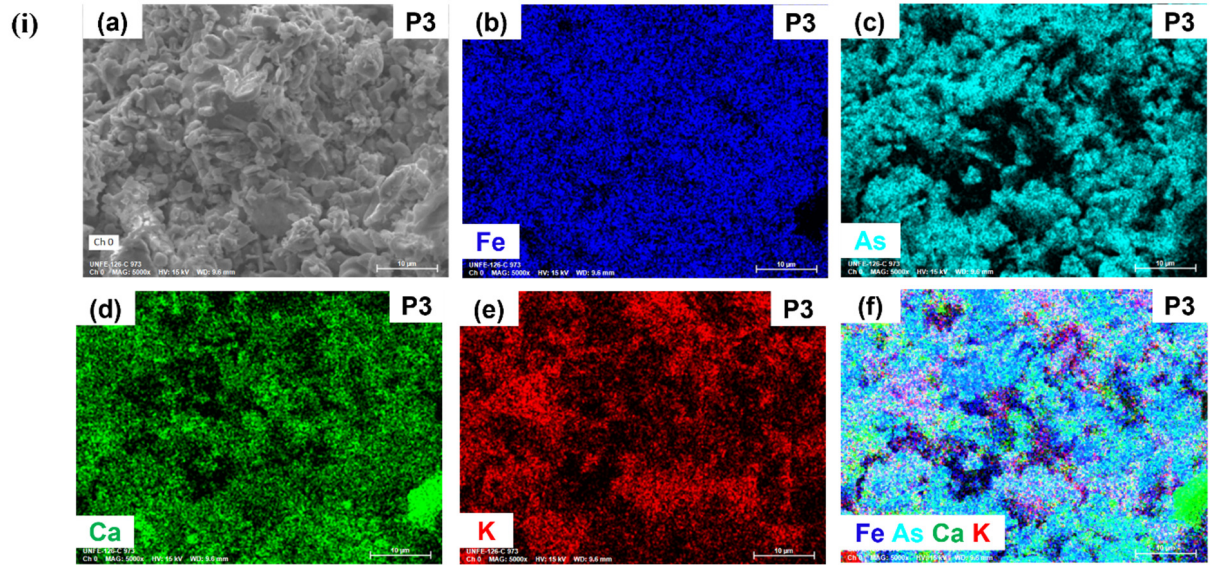

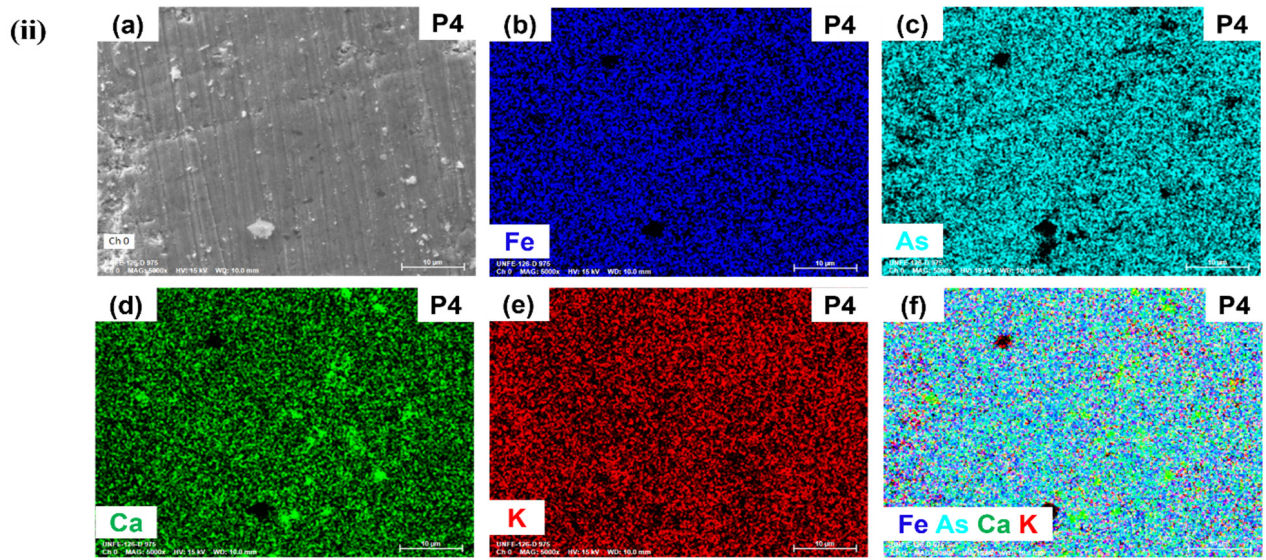

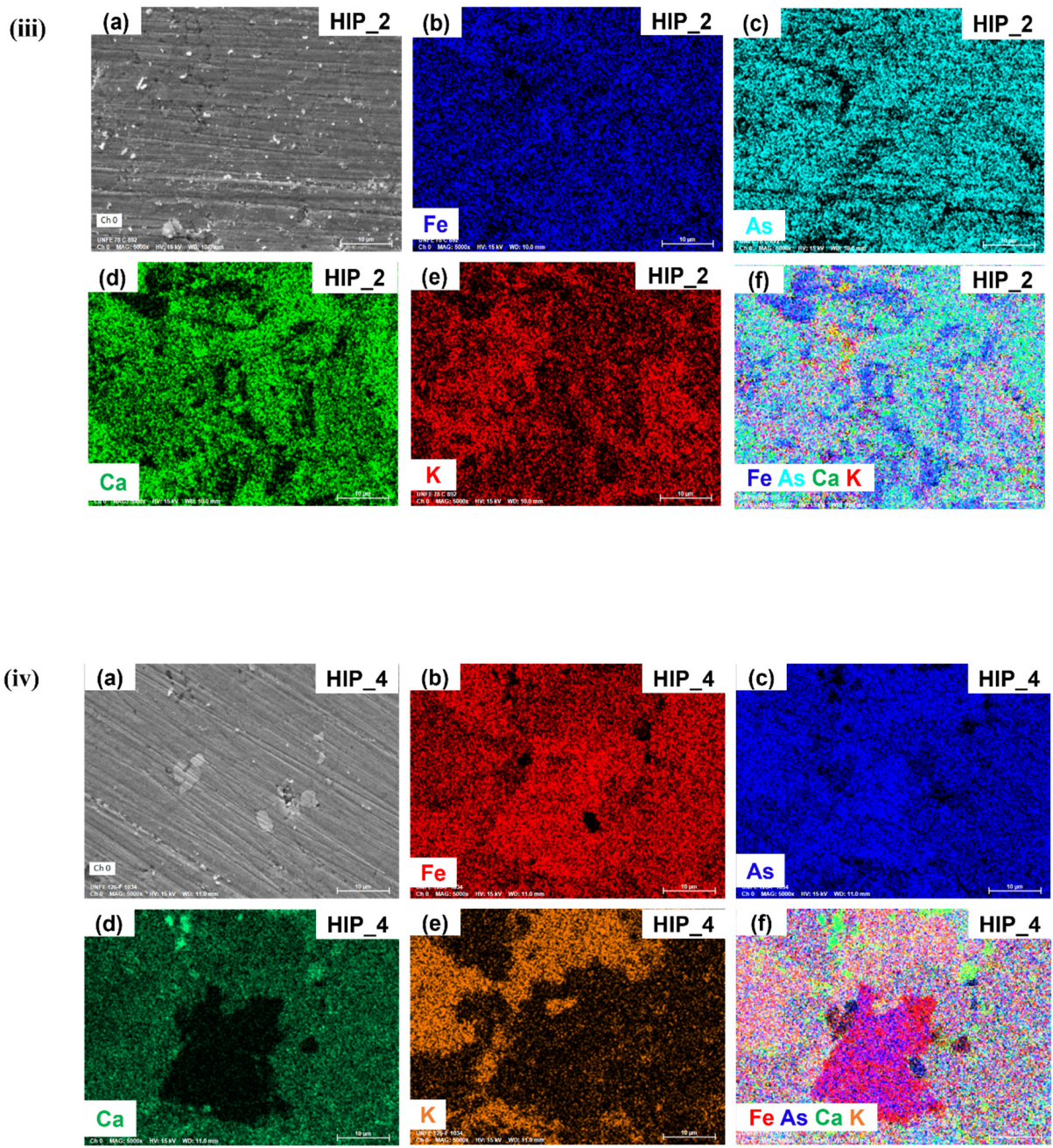

**Figure S2 :** Elemental mapping for various CaKFe$_4$As$_4$ samples: **(i)** P3 **(ii)** P4 **(iii)** HIP_2 **(iv)** HIP_4.

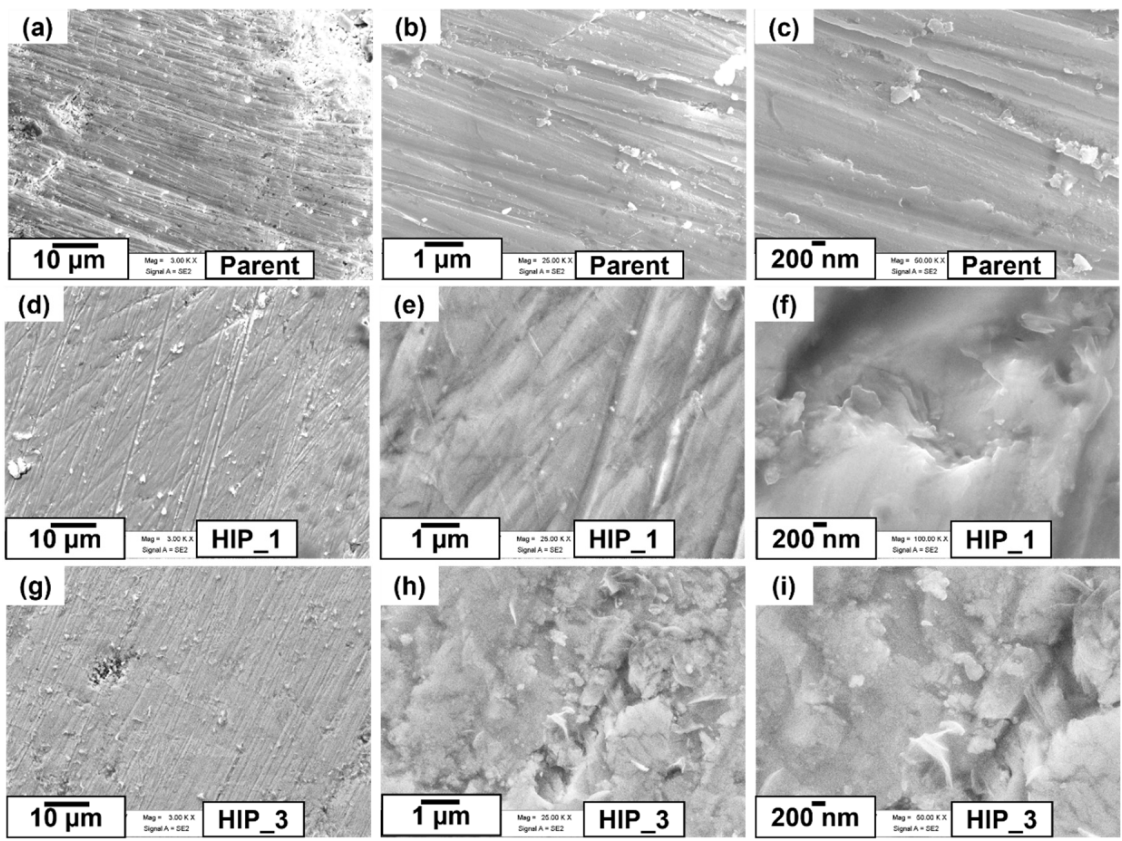

**Figure S3:** Secondary electrons (SE) images at different magnifications for the CaKFe$_4$As$_4$ bulks: **(a)-(c)** Parent, **(d)-(f)** HIP_1, and **(g)-(i)** HIP_3.

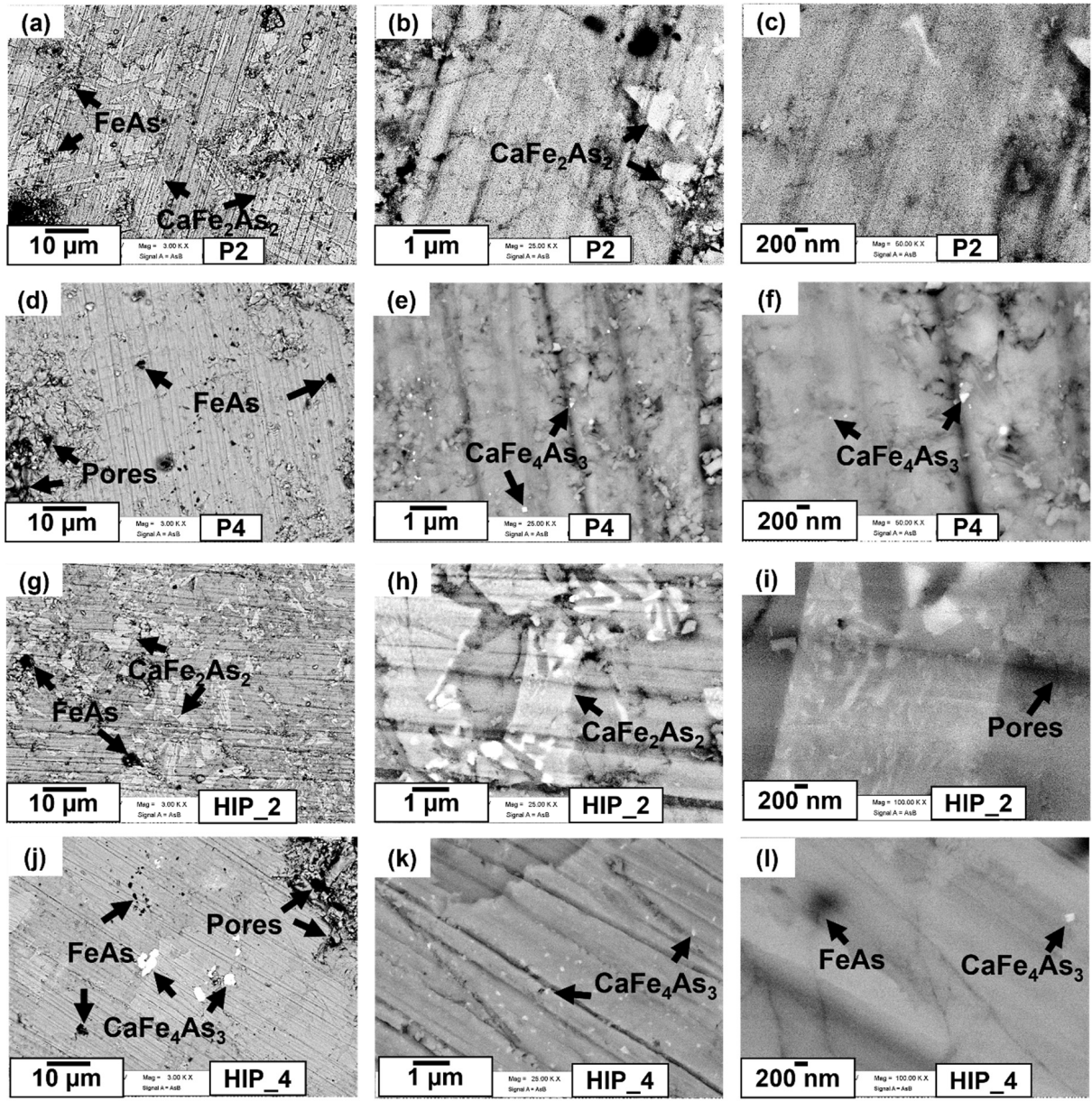

**Figure S4:** Backscattered Electrons (BSE: AsB) images at different magnifications for the CaKFe$_4$As$_4$ bulks: **(a)-(c)** P2, **(d)-(f)** P4, **(g)-(i)** HIP_2 and **(j)-(l)** HIP_4. The white contrast could be either CaFe$_2$As$_2$ or CaFe$_4$As$_3$ phase.

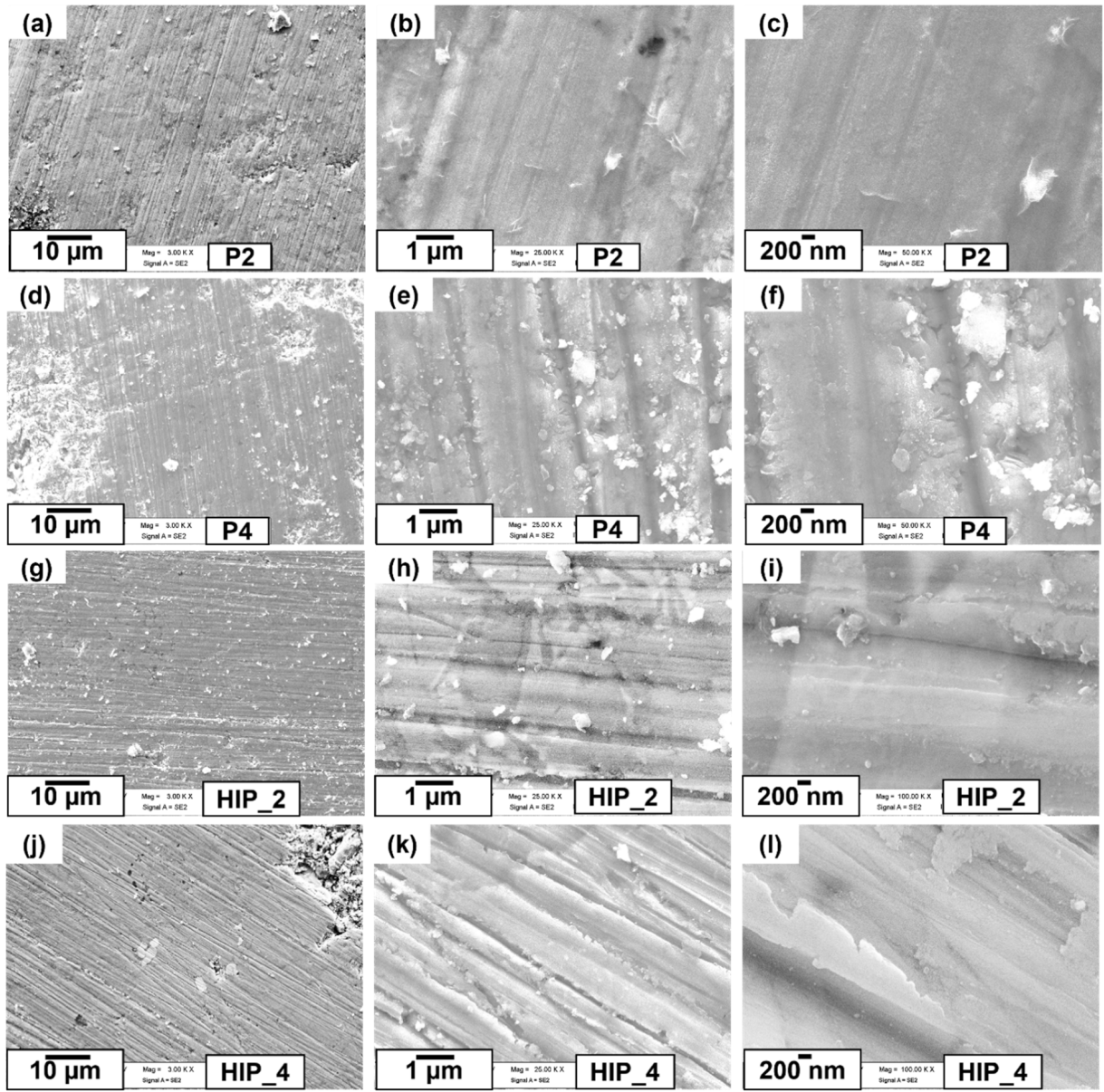

**Figure S5:** Secondary electrons (SE) images at different magnifications for the $CaKFe_4As_4$ bulks: **(a)-(c)** P2, **(d)-(f)** P4, **(g)-(i)** HIP_2 and **(j)-(l)** HIP_4. The white contrast could be either $CaFe_2As_2$ or $CaFe_4As_3$ phase.

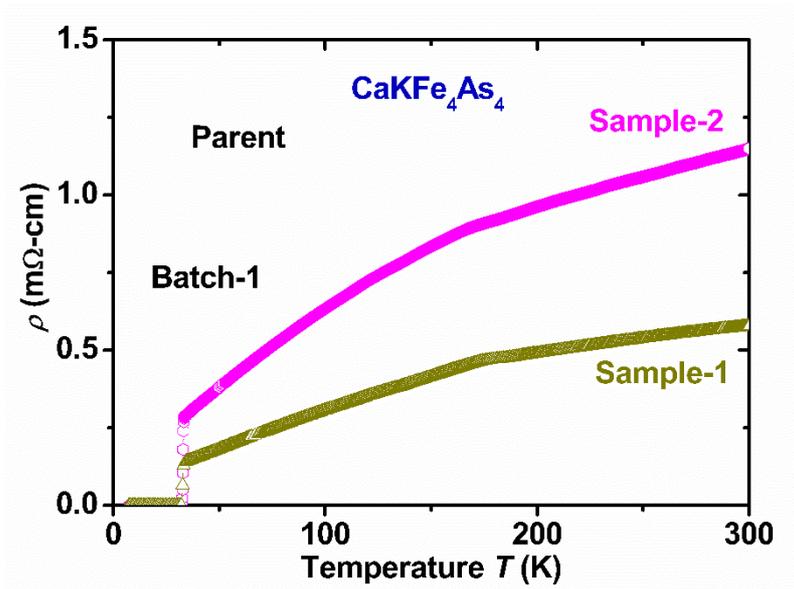

**Fig. S6**: The variation for resistivity ($\rho$) with temperature for two rectangular pieces (Sample-1 and Sample-2) of the parent sample from the same batch (**Batch-1).**

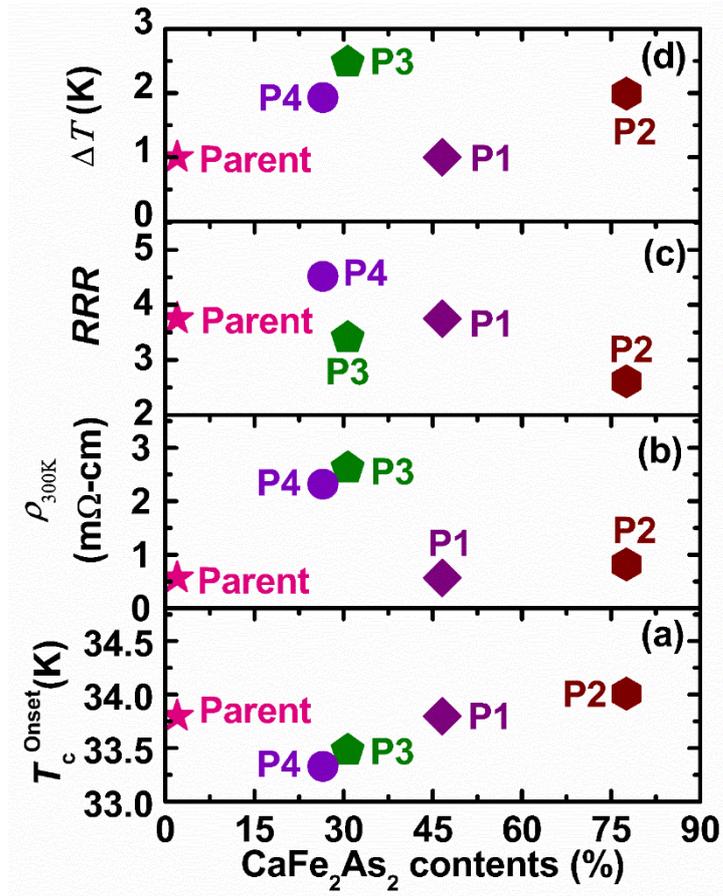

**Fig. S7**: The variation of **(a)** the onset transition temperature $T_c^{onset}$, **(b)** the room temperature resistivity ($\rho_{300K}$), **(c)** Residual resistivity ration ($RRR = \rho_{300\,K} / \rho_{40\,K}$) and **(d)** transition width ($\Delta T$) with respect to an amount (%) of the impurity $CaFe_2As_2$ phase for various $CaKFe_4As_4$ bulks prepared by CSP.

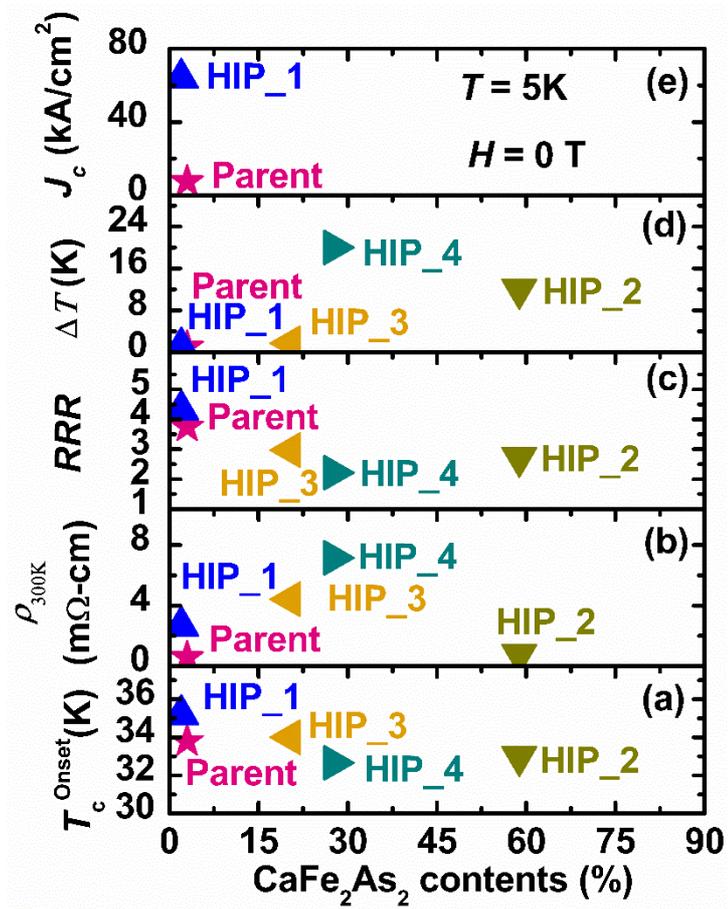

**Fig. S8**: The variation of **(a)** the onset transition temperature $T_c^{onset}$, **(b)** the room temperature resistivity ($\rho_{300K}$), **(c)** Residual resistivity ration ($RRR = \rho_{300\,K} / \rho_{40\,K}$), **(d)** transition width ($\Delta T$) and **(e)** the critical current density ($J_c$) at 5 K and self-field with respect to an amount of the impurity $CaFe_2As_2$ phase for $CaKFe_4As_4$ bulks prepared by HP-HTS method with the "parent" sample prepared by CSP method.